\documentclass{ws-ijmpa}
   \newcommand{\ba}{\begin{eqnarray}}
   \newcommand{\ea}{\end{eqnarray}}
   \newcommand{\be}{\begin{equation}}
   \newcommand{\ee}{\end{equation}}

   \newcommand{\AmS}{{\protect\the\textfont2%
  A\kern-.1667em\lower.5ex\hbox{M}\kern-.125emS}}

\begin{document}

\markboth{J.B. Zhang, D.B. Leinweber, K.F. Liu and A.G. Willams} {Scaling of nonperturbative  
renormalization of composite operators with overlap fermions}

%
\catchline{}{}{}{}{}
%

\title{SCALING OF NONPERTURBATIVE RENORMALIZATION OF \\
COMPOSITE OPERATORS WITH OVERLAP FERMIONS
}

\author{\footnotesize J.B. ZHANG$^{1,2}$, D.B. LEINWEBER$^2$, and
        A.G. WILLIAMS$^2$  
}

\address{$^1$ ZIMP and Department of Physics, \\ 
        Zhejiang University, Hangzhou 310027, P. R. China}

\address{$^2$ Department of Physics and  \\
        Special Research Centre for the Subatomic Structure of Matter,\\
        University of Adelaide, SA 5005, Australia 
}


\maketitle


\begin{abstract}

We compute non-perturbatively the renormalization constants of composite operators
for overlap fermions by using the regularization independent scheme.
The scaling behavior of the renormalization constants is investigated using the data from
three lattices with similar physical volumes and different lattice spacings. 
The approach of the renormalization constants to the continuum limit is explored.

\keywords{lattice QCD; scaling, non-perturbative renormalization; overlap fermion.} 
\end{abstract}

\maketitle

\section{Introduction}

 Following previous papers~\cite{NPR1,NPR2} in which we computed non-perturbatively 
the renormalization constants of composite operators with overlap fermions in quenched QCD,
this paper will study the scaling behavior of the renormalization constants.
 
We adopt the non-perturbative renormalization method which was introduced by      
Martinelli ${\it et~al.}$~\cite{NPM}.
The method allows a full non-perturbative computation of the matrix elements
of composite operators in the Regularization Independent (RI) scheme~\cite{NPM,Tblum1}
(it is called the ${\rm RI^\prime}$ scheme by Chetyrkin~\cite{Chetyrkin:20004l}). 
The matching between the ${\rm RI}$ scheme and $\overline{\rm MS}$, which is
intrinsically perturbative, is computed using continuum perturbation
theory, which is well behaved.

The overlap fermion~\cite{neub1,neub2} was proposed by Narayanan and 
Neuberger to evade the so called ``no-go" theorem~\cite{no-go}.
The action in the massless
limit preserves a lattice form of chiral symmetry
even at finite lattice spacing and volume~\cite{neub2,luscher}.
The use of the
overlap action entails many theoretical advantages: it has
no additive mass renormalization, there are no order $a$ artifacts,
and it has very good scaling~\cite{kent1}.

\section{Non-perturbative renormalization method}
\label{sec:NPRM}

 The renormalized operator $O(\mu)$ is related to the bare operator, $O(a)$, calculated on the lattice via
\begin{equation}
O(\mu) = Z_O(\mu a, g(a))~O(a) \, ,
\end{equation}
In this work, we will consider the fermion operators
\begin{equation}
O_{\Gamma}(x) = \bar{\psi}(x) \Gamma \psi(x) \, ,
\end{equation}
where $\Gamma$ are the Dirac gamma matrices
$
\Gamma \in \left\{ 1 , \gamma_\mu , \gamma_5, \gamma_\mu
\gamma_5, \sigma_{\mu \nu} \right\}$ 
and the corresponding notations will be \{ S, V, P, A, T \} respectively.

The renormalization condition is imposed directly on  the three-point vertex function $\Gamma_O(pa)$,
which is calculated in a fixed gauge, {\it e.g.}, Landau gauge in our case, 
at a momentum scale $p^2 = \mu^2$
\begin{equation}    \label{eq:ren_con}
\Gamma_{O, ren}(pa)|_{p^2 = \mu^2} = \frac{Z_O(\mu a, g(a))}{Z_{\psi}(\mu a, g(a))}
\Gamma_O(pa)|_{p^2 = \mu^2} = 1  \, .
\end{equation}
Here $Z_{\psi}$ is the field or wave-function renormalization constant, $\Psi_{\rm ren} = Z_{\psi}^{1/2} \Psi$.

It is apparent that we can only get the ratio of the renormalization 
constant $Z_O$ for the operator $O$ and the wave-function renormalization constant $Z_\psi$, from the 
renormalization condition of Eq.~(\ref{eq:ren_con}). 
In order to obtain the renormalization constant $Z_O$ for the operator $O$, one needs
to know $Z_{\psi}$ first. 
In this work, we will obtain $Z_{\psi}$ directly from the quark propagator. It
can be defined from the Ward Identity (WI) as~\cite{NPM}
\begin{equation}
Z'_\psi=\left.-i\frac{1}{12}\frac{{\rm{Tr}} \sum_{\mu=1,4}
\gamma_\mu (p_\mu a)S(pa)^{-1}}
{4\sum_{\mu=1,4}(p_\mu a)^2}\right|_{p^2=\mu^2}\; ,
\label{eq:Z_q'_WI}
\end{equation}
which, in Landau gauge, differs from $Z_\psi$ by a finite term of order
$\alpha_s^2$. The matching coefficients have been computed using continuum
perturbation theory~\cite{Franco:1998bm}.

\section{Numerical details}
\label{sec:numerical}

We work on three lattices, each with a different lattice
spacing, $a$, but having similar physical volume. Lattice parameters are summarized in
Table~1. The quenched gauge configurations are  
created using a tadpole improved plaquette plus rectangle
(L\"{u}scher-Weisz) gauge action and gauge fixed to the Landau gauge using a
Conjugate Gradient Fourier Acceleration algorithm.
                                                                                                                                             
\begin{table}[pt]
\tbl{Lattice parameters.}
{\begin{tabular}{@{}ccccccc@{}} \toprule
Action & Size & $N_{\rm{Samp}}$ & $\beta$ &$a$ (fm) & $u_{0}$
 & Physical Volume (fm$^4$)\\ \colrule
Improved       & $16^3\times{32}$ & 500 & 4.80 & 0.093  & 0.89650 &
$1.5^3\times{3.0}$ \\
Improved       & $12^3\times{24}$ & 500 & 4.60 & 0.123  & 0.88888 &
$1.5^3\times{3.0}$ \\
Improved       & $8^3\times{16}$  & 500 & 4.286& 0.190  & 0.87209 &
$1.5^3\times{3.0}$ \\   \botrule
\end{tabular}}
\label{parameter}
\end{table}

The overlap-Dirac operator we use is 
\begin{equation}  \label{D_m}
D(m_q) = \rho+\frac{m_q}{2} + (\rho-\frac{m_q}{2})\gamma_5\epsilon(H).
\end{equation}
Where $\rho$ is regulator mass and $m_q$ is bare quark mass, and $\epsilon(H)$ is the matrix sign function 
of an Hermitian operator $H$ = $\gamma_5 D_W$. 
We use the tadple improved Wilson kernel $D_W$ in the overlap operator, and  $\kappa=0.19163$ is  used for the
regulator mass $\rho$ for all three lattices.  We calculate the 
overlap quark propagator for 15 bare quark masses on three lattices, 
they are $53$, $59$, $71$, $83$, $94$,$106$,
$124$, $142$, $177$, $212$, $266$,
$554$, $442$, $531$, and $620$ ~MeV respectively

Our calculation begins with the evaluation of the
inverse of the overlap-Dirac operator.  
After we calculate the quark propagator in coordinate space for each configuration,
we use the Landau gauge fixing transformation matrix to rotate the quark propagator
to Landau gauge. Then the discrete Fourier transformation is used to obtain the quark propagator
in momentum space. 
Afterward, we calculate five projected vertex functions $\Gamma_O(pa)$ for each bare quark mass, and 
extrapolate to the chiral limit.
These projected vertex functions $\Gamma_O(pa)$ are in general dependent on $(pa)^2$. The dependence may come from
two sources.  One is  from the usual running of the renormalization constant in the ${\rm RI}$ scheme. The other
is from possible $(pa)^2$ errors and we need to remove this. 
In order to confront experiment, it is preferable to quote the final
results in the $\overline{
\rm MS}$ scheme at a certain scale. One needs to transform
the results in the ${\rm RI}$ scheme to the $\overline{
\rm MS}$ scheme. The detailed analysis can be found in Refs.~\cite{NPR1,NPR2}.

\section{Scaling behaviors}
\label{sec:scaling}

\begin{figure}[ht]
\label{fig3lz}
\begin{tabular}{lr}
{\epsfig{angle=90,figure=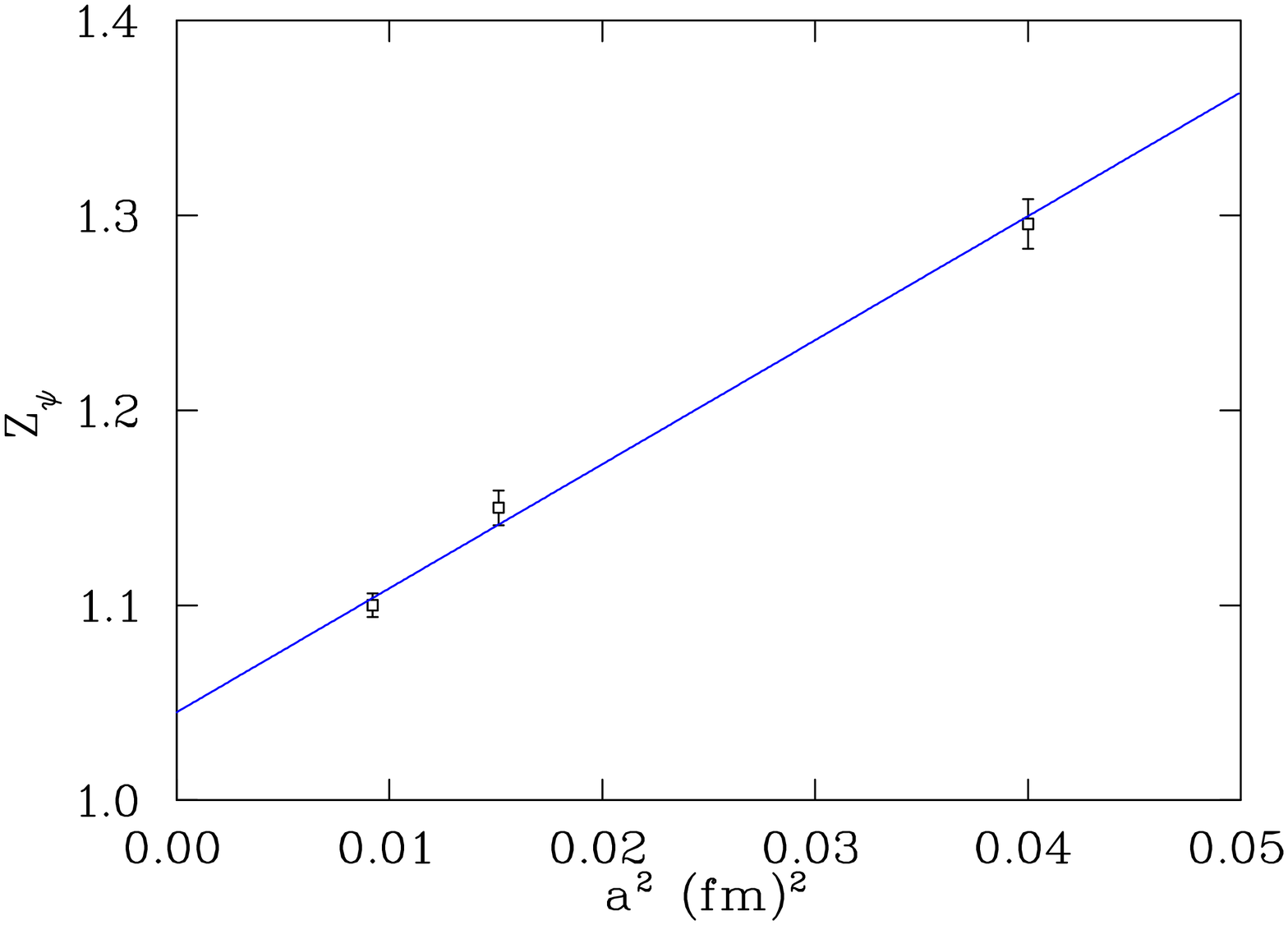,height=4cm} }
{\epsfig{angle=90,figure=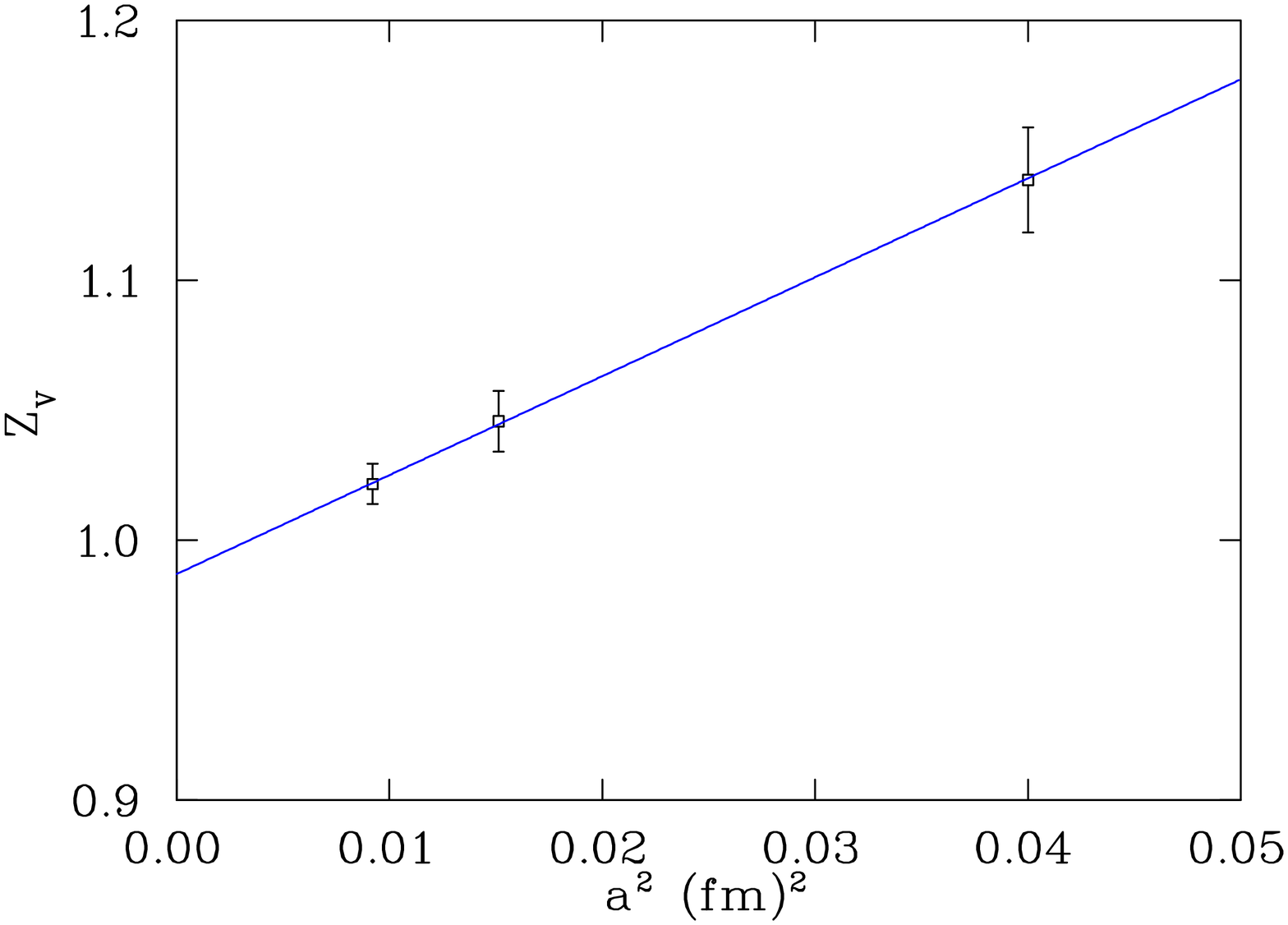,height=4cm} } \cr
{\epsfig{angle=90,figure=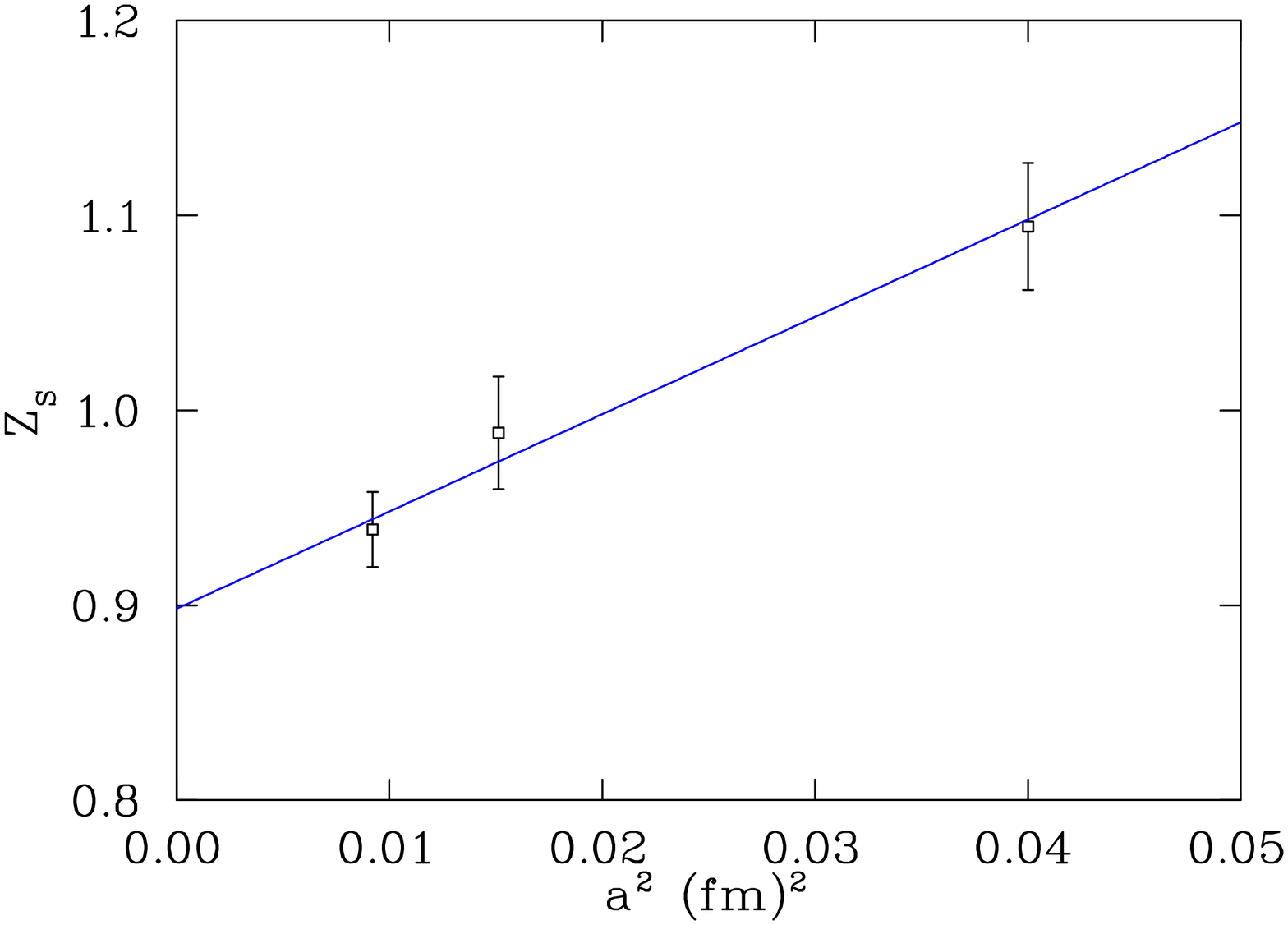,height=4cm} }
{\epsfig{angle=90,figure=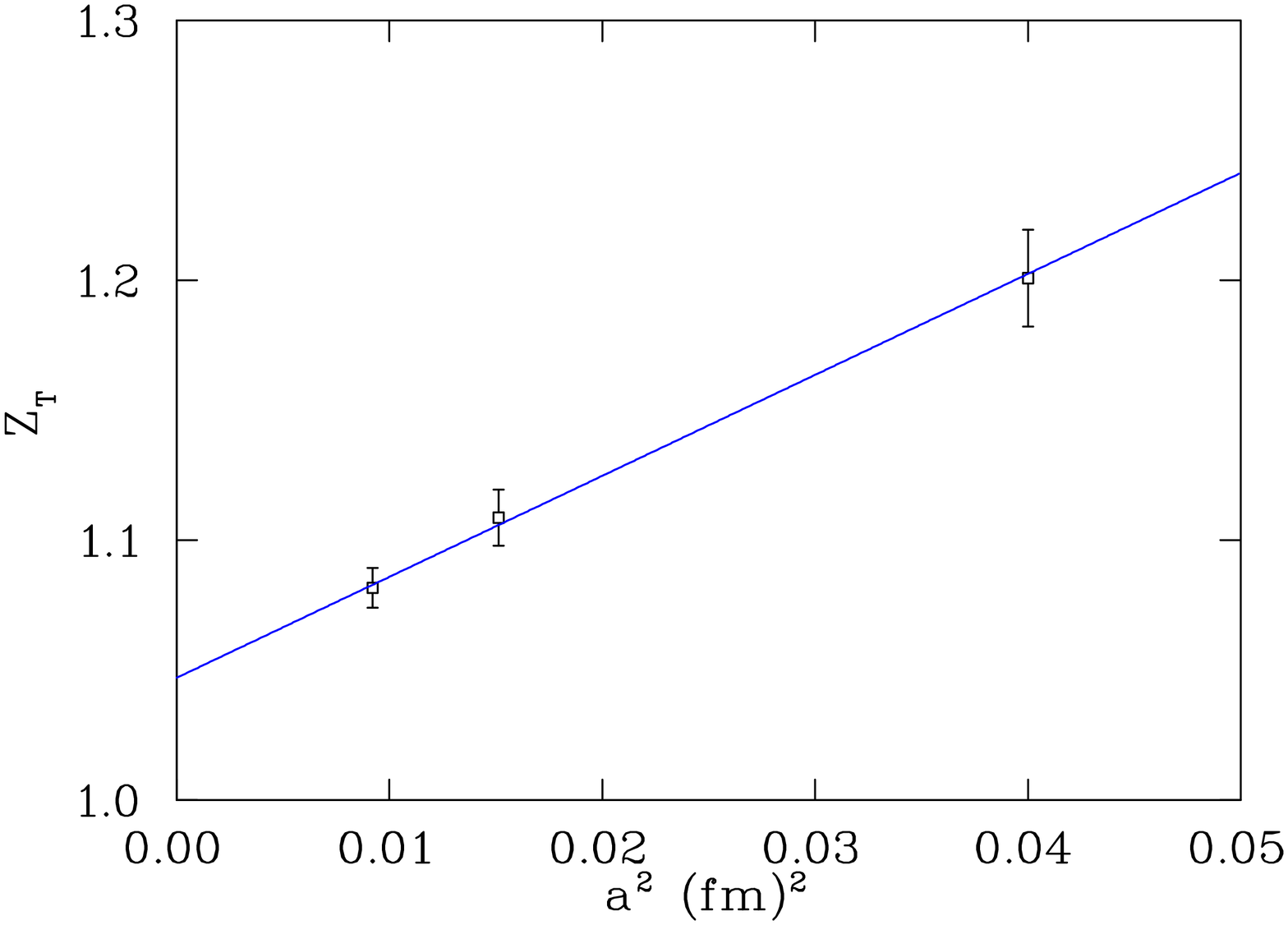,height=4cm} }
\end{tabular}
\caption{The renormalization constants $Z_\psi$, $Z_A$, $Z_S$, and $Z_T$ in ${\overline{\rm MS}}$ at 2.0 GeV against 
the square of the lattice spacing $a$. The straight line is the linear fit $Z_O$ = $c_1+c_2 a^2$.}
\end{figure}

  We work on three lattices with similar physical volumes and different spacings $a$ to investigate the
the scaling behavior of the renormalization constants. 
Here we compare the results of renormalization 
constants  $Z_\psi$, $Z_V$, $Z_S$ and $Z_T$ on the different lattices in the ${\overline{\rm MS}}$ scheme at 2.0 GeV. 
Fig.~\ref{fig3lz} shows the four renormalization constants  $Z_\psi$, $Z_V$, $Z_S$ and $Z_T$ against
the square of the lattice spacing $a$. Because overlap fermions are free of $O(a^2)$ errors, the leading term must be
proportional to $a^2$.  We use a simple linear fit $Z_O$ = $c_1$ +
$c_2$ $a^2$, and take $c_1$ as the value of the renormalization constant $Z_O$ in the continuum limit.
The numerical values of calculated renormalization constants in the continuum limit in the ${\overline{\rm MS}}$ and RI
schemes at 2.0 GeV are displayed in Table~2.

  In the continuum, the vector current is conserved, so $Z_V$ should be equal to one.  For the axial vector
current, due to the PCAC relation, it will be conserved at large momenta, where the anomaly has no effect.
Our result for $Z_V$ and $Z_A$ in Table~2 compares favorably with 1.

\begin{table}[pb]
\label{zresult_c}
\tbl{Results for $Z$ in the continuum limit.} 
{\begin{tabular}{@{}cccc@{}} \toprule
$Z-factor $  & ${\rm RI}$ scheme at 2 GeV & $ {\overline{
\rm MS}}$ scheme at 2 GeV       \\  \colrule
$Z_\psi$     & 1.059$\pm$0.008  & 1.045$\pm$0.008    \\
$Z_V$($Z_A$) & 0.987$\pm$0.011  & 0.987$\pm$0.011    \\
$Z_S$($Z_P$) & 0.765$\pm$0.021  & 0.898$\pm$0.025    \\
$Z_T$        & 1.069$\pm$0.010  & 1.047$\pm$0.010    \\  \botrule
\end{tabular}}
\end{table}



\end{document}